\documentclass[prl,twocolumn,english,showpacs]{revtex4-1}
\usepackage{graphicx}
\usepackage{amssymb}
\usepackage{braket}
\usepackage[rightcaption]{sidecap}%
\graphicspath{{figures/}}

\begin{document}

\title{Relaxation dynamics of a Fermi gas in an optical superlattice}

\author{D. Pertot$^{1,2}$, A. Sheikhan$^3$, E. Cocchi$^{1,2}$, L. A. Miller$^{1,2}$, J. E. Bohn$^{2}$,  M. Koschorreck$^{1,2}$, M. K{\"o}hl$^{1,2}$, and C. Kollath$^3$}

\affiliation{$^1$Physikalisches Institut, University of Bonn, Wegelerstra\ss e 8, 53115 Bonn, Germany\\
$^2$Cavendish\,Laboratory,\,University~of Cambridge, JJ Thomson Avenue, Cambridge CB30HE, United Kingdom\\
$^3$HISKP, University of Bonn, Nussallee 14-16, 53115 Bonn, Germany}

\begin{abstract}
This paper comprises an experimental and theoretical investigation of the time evolution of a Fermi gas following fast and slow quenches of a one-dimensional optical double-well superlattice potential. We investigate both the local tunneling in the connected double wells and the global dynamics towards a steady state. The local observables in the steady-state resemble those of an equilibrium state, whereas the global properties indicate a strong non-equilibrium situation. 
\end{abstract}

\pacs{03.75.Ss %Fermi gases
05.30.Fk %Fermion many-body systems
68.65.-k %Low-dimensional, mesoscopic, nanoscale and other related systems: structure and nonelectronic properties
}

\date{\today}
\maketitle

One of the grand goals of condensed matter physics is the dynamic control of the quantum properties of materials. The target is the creation of new functionality as well as the dynamic change in between different phases. Recently, great experimental progress has been made in this direction~\cite{BasovHaule2011}. Examples are photo-induced phase transitions as in the insulator-to-metal transitions, fast demagnetization and magnetization control, or the generation of Josephson plasmons (which are typically present in superconducting phases) in non-superconducting striped-order cuprates ~\cite{FaustiCavalleri2011}. However, the underlying dynamical mechanisms are often far from being understood, since  many different processes in the form of phononic relaxation channels are present in  the solid state. 

Recent progress towards understanding non-equilibrium physics in the solid state has been complemented by advances in ultracold atomic physics. Atomic quantum gases are excellent systems in which to study quantum dynamics \cite{Bloch2008,Lewenstein2007} because of the clean preparation, wide tunability and long coherence times. In particular, different dynamical processes can be well separated and timescales for the real-time detection of dynamics are experimentally favourable. The fastest relevant energy scale, the Fermi energy, is on the order of 10kHz and can be well resolved. So far, experimental progress in the probing and understanding of dynamical phenomena in optical lattices has been mostly focused on bosonic systems where, for example, interaction quenches have been studied \cite{Greiner2002b,Sebbystrabley2007,Foelling2007,Cheneau2012,Trotzky2012,Hung2010,Sherson2010,Bakr2009,Chen2011,Ronzheimer2013}.
With fermionic gases in optical lattices, interaction quenches \cite{Strohmaier2007,Hackermueller2010,Schneider2012,Tarruell2012} and spin dynamics have been investigated \cite{Klauser2012,Koschorreck2013}.

In this paper we investigate experimentally and theoretically the time evolution of a Fermi gas in a bichromatic optical lattice which is imposed along one dimension of the gas, following a sudden or a slow quench of the lattice potential. We initially prepare a patterned density distribution in the optical lattice and measure the relaxation of both local and non-local observables towards a steady-state.

Experimentally \cite{Frohlich2011}, we start from a non-interacting degenerate gas of $N=1.2 (2) \cdot 10^5$ $^{40}$K atoms in each of the two lowest hyperfine spin states $\Ket{F=9/2,m_F=-9/2}$ and $\ket{F=9/2,m_F=-7/2}$ in a dipole trap with trapping frequencies $\omega_x= 2\pi \times 38$\,Hz, $\omega_y=2\pi \times 66$\,Hz and $\omega_z=2 \pi \times 376$\,Hz. The scattering length between the two components is set to zero by means of a Feshbach resonance. The gas has a temperature of $(0.3\pm 0.1)\,T_F$ where $k_BT_F=E_F$ is the Fermi energy of the gas. We then apply a bichromatic optical lattice along the x-axis which consists of a short-period (``green") lattice of wavelength $\lambda_G=532$\,nm and a long-period (``red") lattice of wavelength $\lambda_R=1064$\,nm (see supplementary material). The bichromatic lattice potential at the location of the atoms is given by $V(x)=V_G \cos^2(k_G x-\phi) - V_R \cos^2(k_R x)$,
where $V_G$ ($V_R$) is the potential depth of the green (red) lattice and $k_{R/G}=2 \pi/\lambda_{R/G}$. The periods of the two lattices differ by a factor of two, up to a small, controlled and adjustable difference. This difference results in a relative phase $\phi$ between the two lattices \cite{Foelling2007} and permits the generation of distinct double-well configurations. For $\phi=0$ the double-well configuration is symmetric and for  $\phi \neq 0$ there is an energy offset between adjacent wells (see Figure 1). We use this control over the lattice potential for a density-patterned loading of the atomic ensemble and for the subsequent discrimination of the density in even and odd wells during measurement \cite{Sebbystrabley2007,Foelling2007,Tarruell2012}.

\begin{figure}
\includegraphics[width=\columnwidth,clip=true]{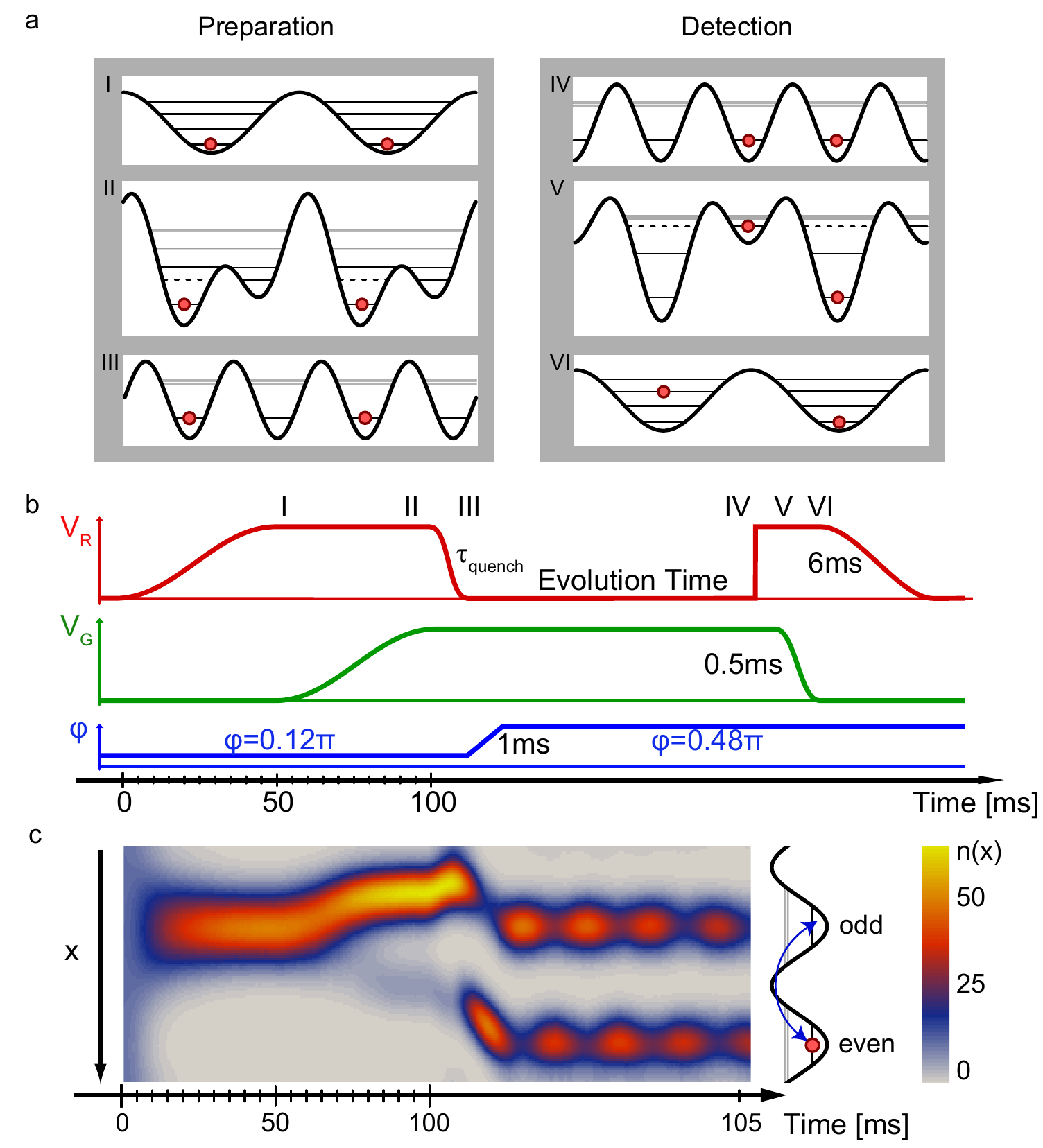}
 \caption{(Color online) \textbf{a)} Sketch of the preparation and of the detection of the local odd-even density imbalance according to the lattice ramp sequence shown in \textbf{b)}. \textbf{c)} Simulated time-evolution of the density distribution in the central double well integrated along the $y-$ and $z$--directions.}
\label{fig1}
\end{figure}

The ramp sequence for the optical lattice potentials and the resulting lowest eigenenergies of the central double-wells are sketched in Figs.~\ref{fig1}a, b. The Fermi gas is initially loaded into the red lattice by ramping up the depth to $V_R=48 E_{rec,R}$ in 50\,ms, where $E_{rec,R/G}=\hbar^2k_{R/G}^2/(2 m)$ is the recoil energy and $m$ is the mass.  This ramp is chosen slow enough such that predominantly states in the lowest Bloch band are occupied, which are well localized within the red lattice wells (Fig.~1a,I). Subsequently, the green lattice is ramped to a variable depth $V_G$, also within 50 ms. This splits each well of the red lattice almost adiabatically into two wells. For the chosen value $\phi= 0.12(2) \pi$, the states of the lowest energy band lie within the odd-numbered wells of the double-well structure (see Fig. 1a, II). In Fig.~1c we show the simulated time-evolution of the density along the $x$--direction (integrated along the $y$-- and $z$--directions) for one double-well. After the preparation ($t=100$\,ms) a high occupation in the deep (odd) wells and a very low occupation in the shallow (even) wells is generated. We have performed simulations of the discretized version of the model for the entire loading procedure for systems of up to $250$ double wells with $2\times10^5$ atoms. 

In the first set of experiments we perform a rapid quench from the asymmetric double-well lattice (Fig.~1a, II) to the plain green lattice (Fig.~1a, III) where no energy offset between the sites exists, and we monitor the evolution dynamics. The initially strongly-imbalanced, patterned distribution of fermions, which predominantly occupy every other green lattice well, is observed to oscillate between even and odd wells and relaxes over longer times to a balanced distribution. To initiate the quench, we quickly ramp down the red lattice to zero within $\tau_\mathrm{quench}=0.5$\,ms, which is shorter than or comparable to the tunneling times depending on the green lattice depth $V_G$. Subsequently, we shift the superlattice phase offset to $\phi = 0.48(2) \pi$ in preparation for detection. The system then evolves during the variable free evolution time in the green lattice and  we detect the local density in odd and even wells along the $x$--direction (integrated along $y$ and $z$ direction). This is done by projecting the system back into an arrangement of double-wells and subsequently performing adiabatic band-mapping (see the supplementary material). By this detection process the occupation in the odd (even) sites is mapped onto the 1st (3rd and 4th) Brillouin zone \cite{Sebbystrabley2007,Foelling2007}, see Fig. 1a, IV--VI. 

In order to quantify the dynamics, we define the odd-even contrast $C_{exp}=\frac{N_1-(N_3+N_4)}{N_1+N_3+N_4}$ where $N_i$ is the number of atoms in the $i$-th Brillouin zone, i.e. the occupation of the quasimomentum states in the interval $2(i-1)\pi/\lambda_R\leq |q|< 2i\pi/\lambda_R$. The odd-even contrast yields $C_{exp}=1$ if all atoms are in the odd wells, $C_{exp}=-1$ if they are all in the even wells, and $C_{exp}=0$ if the distribution is balanced. Fig. 2a shows the tunneling dynamics in the odd-even contrast for a green lattice of depth $V_G= (11.7\pm 0.9)\,E_{rec,G}$ which displays a damped oscillation. 

A first insight into the dynamics can be gained by considering a one-dimensional tight-binding model with nearest-neighbour tunneling of amplitude $J$. Starting with an initial state with only odd sites occupied by one fermion, the local densities are predicted to evolve according to $\langle n_j(t)\rangle=1/2-(-1)^j J_0(4Jt/\hbar)/2$ as also found for the bosonic case \cite{Flesch2008}. Here $J_0(\cdot)$ is the zeroth-order Bessel function and $j$ denotes the lattice site. We have solved the one-dimensional tight-binding model for any initial imbalance prepared in the experiment (see Fig.~2d). In contrast to this simple model, in the present experiment the fermions occupy many harmonic energy levels along the $y$-- and $z$-- directions. 

We fit our data for $C_{exp}$ with the function $A\sin(2 \pi t/T+\Phi)/(t/T)^\alpha+\textrm{const.}$ from which we extract the oscillation period $T$ and the exponent $\alpha$ characterizing the damping of the oscillations (Fig.~2a). The finite offset of the contrast at long times is due to the experimental detection procedure and corresponds approximately to a balanced situation \cite{Sheikhan2014}. The value $\alpha=1/2$ would correspond to the asymptotic behaviour of the Bessel function solution for $4Jt/\hbar \gg 1$. The measured period of the oscillations agrees excellently with the prediction for the period $T=h/(4J_G)$, which we have confirmed for different lattice depths $V_G$ (Figure 2b). In our experimental data the decay of the coherence of the local densities occurs with an exponent $\alpha \approx 2$  (Figure~2c) which is faster than the Bessel-function prediction, a behaviour which has previously been observed for interacting bosonic atoms in one dimension \cite{Trotzky2012}.

In order to better understand the experimental data, we have performed numerical simulations of the trapped system at zero temperature taking the $\sin^2(x)$ potential of the lattice explicitly into account. The results for the local odd-even density imbalance of the continuum model are shown in Fig.~2e. The local density $n_j$ is the continuum density $n({\bf r})$ integrated over the $j$th lattice well along the $x$ direction and along the entire $y$ and $z$ directions. The central wells of the trap approximately follow the evolution of the one-dimensional tight-binding model (Fig.~2d). However, the oscillations of the local odd-even contrast at the boundaries of the cloud, located at double-well indices $l\sim \pm 70$, display a considerable slowing down and Bloch oscillations appear (see Fig. 2e). The oscillating features along the direction of the superlattice are due to the initial preparation. Both the slowing down of the tunneling and the onset of Bloch oscillations are due to the local energy offset between neighbouring wells caused by the harmonic trapping potential. The position of the maximum contrast closely follows the {\it local} Bloch oscillation period $T_B(l)=\frac{8h}{m \omega_x^2 \lambda_R^2 l }$ using the local offsets caused by the trap. The spatially changing period leads to dephasing of the tunneling oscillations and hence to a damped trap-averaged contrast $C=\frac{1}{N}\sum_l (n_{2l+1}-n_{2l})$ (Fig.~2d). The theoretical results for the trap-averaged odd-even contrast  initially show an exponent of 0.5, as expected for a Bessel function, but this becomes larger at longer times. The damping in the experiment appears to be more pronounced than in theory (Fig.~2c). This might stem from the coupling of the different lattice directions by anharmonicity of the lattice potential or finite temperature effects not accounted for in the theoretical simulations.

\begin{figure}
\includegraphics[width=\columnwidth,clip=true]{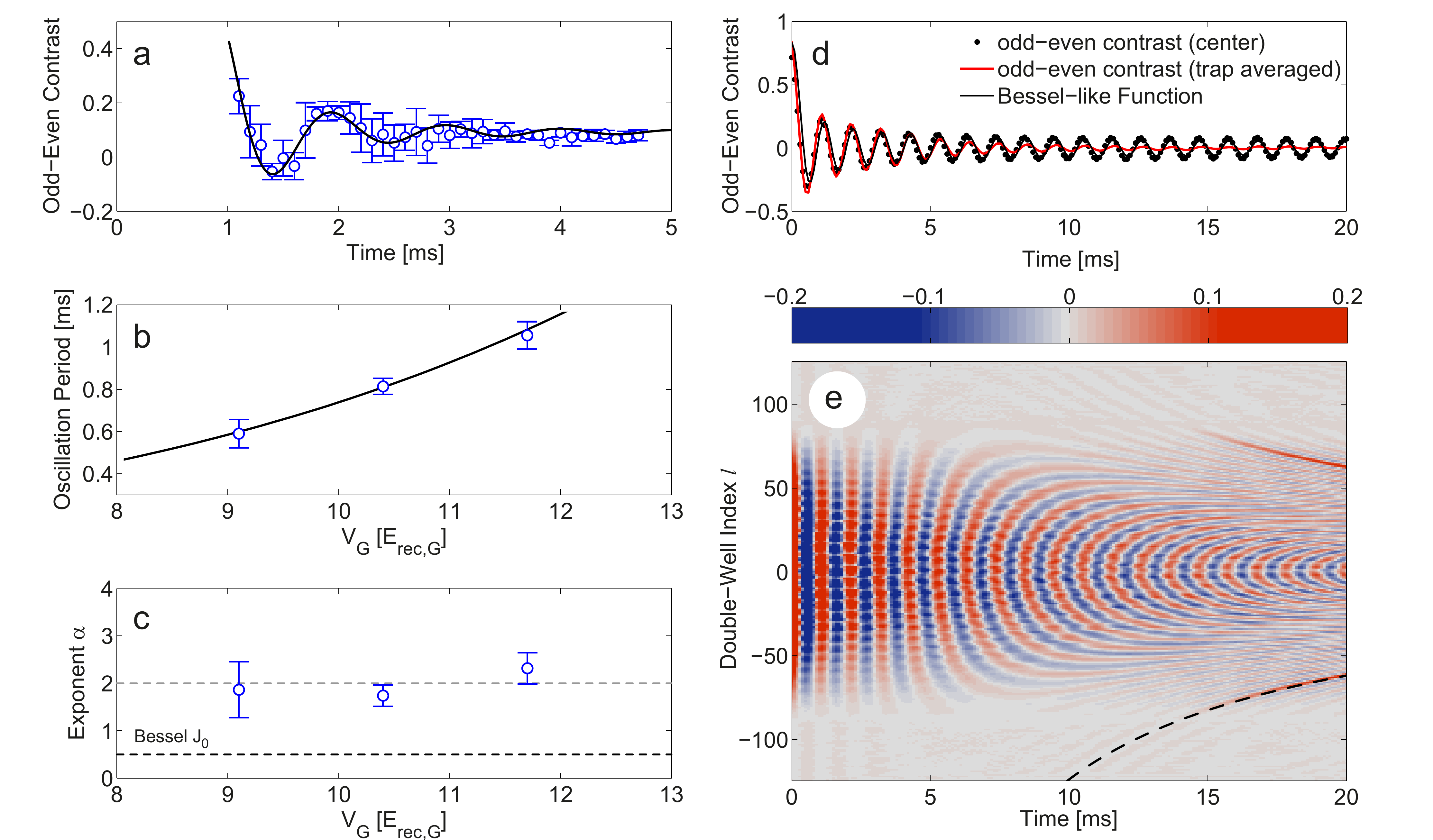}
 \caption{(Color online) \textbf{(a)} Odd-even contrast $C_{exp}$ together with a fit of a damped oscillation for $V_G=11.7(9)\,E_{rec,G}$. Error bars show the statistical error of four averaged data sets.
\textbf{b)} Measured oscillation period vs. lattice depth together with the tight-binding prediction $T=h/(4J_G)$, where $J_G$  is obtained from a  band structure calculation. 
\textbf{c)} Experimental power-law exponent of the damped oscillation. In b) and c) error bars are the fit errors.
\textbf{d)} Theoretical results for the odd-even contrast ($V_G=11.7\,E_{rec,G}$).  Dots show the local results for the center of the trap $c_0=\sum \frac{n_{2l+1}-n_{2l}}{n_{2l+1}+n_{2l}}$, averaged over 10 double wells, which is approximately reproduced by the Bessel-like function with an initial potential offset of $\Delta =4J$ (black line) (see Supplementary material). The red solid line shows the density-averaged contrast $C$ as defined in the text. 
\textbf{e)} Theoretical evolution of the local odd-even density imbalance $(n_{2l+1}-n_{2l})/n_1(t=0)$ in the trap. The black dashed line shows the predicted Bloch period $T_B(l)$.}
\label{fig2}
\end{figure}

To summarize, we have observed that after a rapid quench of the double-well potential, local relaxation towards the balanced situation occurs which corresponds to the expected balanced equilibrium situation. This local relaxation already occurs in a simple integrable model with infinitely many conserved quantities by a dephasing and the relaxation time-scale is shortened by the presence of a trapping potential. The integrability shows up in our theoretical results as a strongly non-thermal distribution of the density on length-scales of the cloud on long time scales.

In a second experimental sequence the global relaxation dynamics to a non-equilibrium state and its dependence on  quenches to different final values are investigated in more detail. For this we perform a slow quench of the superlattice (after a similar initial preparation) by ramping down the red lattice $V_R$ over the time scale $\tau_\mathrm{quench}=6$\,ms to various final lattice depths $V_{R,f}$. 

The dynamics resulting from the slow ramp-down consists of a fast local dynamics, which typically leads to a decrease of the local particle imbalance (as discussed in detail for the rapid quench), and a slower global dynamics which acts to redistribute the overall density profile. The density profile prepared before the quench is typically broader than the corresponding ground state distribution, since only every second lattice well is populated. Choosing different final depths $V_{R,f}$ of the red lattice dictates the magnitude of the small energy offset between neighbouring wells and hence gives us control over the tunnelling dynamics. We measure the time evolution of the global relaxation by detecting the quasimomentum distribution with adiabatic band mapping \cite{Greiner2001b} of the green lattice (2\,ms). The band-mapping technique projects the state onto the occupation of the quasi-momentum states of the homogeneous (untrapped) system. In the absence of a harmonic trap, the quasimomentum distribution should remain stationary after the quench. However, the presence of the harmonic trap couples different quasi-momentum states and thus the probe is largely insensitive to the local tunneling dynamics, but instead probes global scales. In the theoretical simulations we obtain the quasi-momentum distribution by direct projection onto the homogeneous quasi-momenta.

The evolution of the experimental and theoretical quasimomentum occupation is shown in Fig.~3 for final lattice depths $V_{R,f}=0$ (Fig. 3 a,b) and $\simeq 0.1\,E_{rec,R}$ (Fig. 3 c,d). In the green lattice alone without an offset between neighbouring wells $(V_{R,f}=0)$ we initially observe only quasi-momentum states occupied in half of the first Brillouin zone of the green lattice, i.e.,~the quasimomentum states $|q|< \pi/\lambda_G$. Owing to the effects of the harmonic trap, the quasimomentum distribution expands quickly $(\sim $\,ms) into the second region ($\pi/\lambda_G\leq |q|< 2\pi/\lambda_G$) of the Brillouin zone. As we will see later, this is mainly due to the redistribution of atoms in the central region of the trap. After a timescale of approximately $10$\,ms (roughly half the dipole trap frequency), the  Brillouin zone up to $q=2\pi/\lambda_G$ is occupied and for longer times, a fishbone-like structure emerges on the timescale of the trapping period, hints of which are visible in the experimental data. This structure is caused by a breathing mode induced by the lattice ramp which has  approximately the expected period of $36$ms (related to the rescaled trap frequency $\Omega=\pi \omega_x\sqrt{J_G/E_{rec,G}}$). 

For a final red lattice depth of $V_{R,f}=0.1 E_{rec,R}$, the homogeneous lattice exhibits a small bandgap at the wave vector corresponding to half the green Brillouin zone with a magnitude of  $\sim0.03\,E_{rec,R}$ (Fig.~4a). This band gap acts to protect the occupation of the quasimomentum states in the range $|q|\leq \pi/\lambda_G$. In the presence of the harmonic trapping potential, the global band gap is destroyed and therefore an occupation of excited quasimomentum states $\pi/\lambda_G<|q|\leq 2\pi/\lambda_G$ sets in. However, compared to the gapless case $V_{R,f}=0$, the transfer of atoms is still strongly suppressed and as a result the occupation of quasimomentum states with $|q|>\pi/\lambda_G$ is reduced (see Fig. 3 c,d).

\begin{figure}
\includegraphics[width=\columnwidth,clip=true]{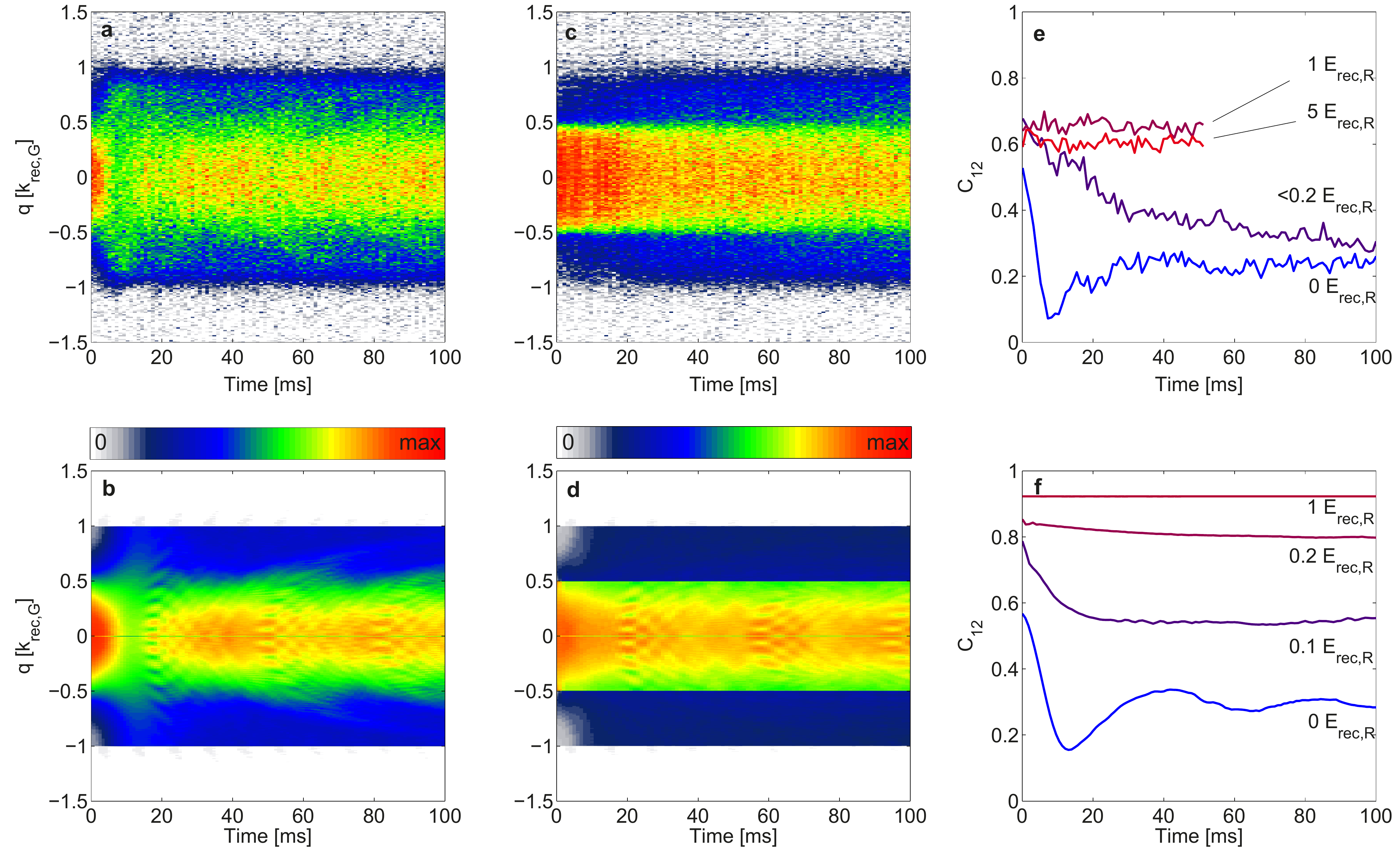}
 \caption{(Color online) Evolution of the quasimomentum distribution after the slow quench for $V_{R,f}=0$ for experimen (a) and theory (b). The middle column show the experimental result for $V_{R,f}<0.2 E_{rec,R}$ (c) compared with the theoretical $V_{R,f}=0.1 E_{rec,R}$ (d). (e,f) Evolution of the quasimomentum contrast $C_{12}$ for different values of $V_{R,f}$. The top row shows experimental data and the bottom row the corresponding theoretical results. The time-axis of the theoretical result at $V_{R,f}=0$ is rescaled by a factor 0.87 in order to better match the experimentally observed features. This factor can stem from the uncertainty of the experimental parameters.  
\label{fig3}}
\end{figure}

In order to quantify the dynamics for different final lattice depths, we define a contrast within the  Brillouin zone  $C_{12}=(N_{|q|<\pi/\lambda_G}-N_{|q|\geq\pi/\lambda_G})/(N_{|q|<\pi/\lambda_G}+N_{|q|\geq\pi/\lambda_G})$. In Figure 3e-f, we show the time evolution for various values of $V_{R,f}$ for a fixed phase offset $\phi=0.12(2) \pi$. For zero red lattice depth, we observe a strongly damped oscillation. This oscillation stems from a breathing mode excited by the global non-equilibrium distribution and results in the fishbone-like structure seen in Fig.~3 a,b. For non-vanishing final lattice depth the resulting asymmetry in the bias of even and odd sites (and the resulting local gap) increasingly prohibits the redistribution of quasimomenta until the quasimomentum distribution becomes frozen for $V_{R,f}\gtrsim 0.5\,E_{rec,R}$.  

\begin{figure}
\includegraphics[width=.6\columnwidth,clip=true]{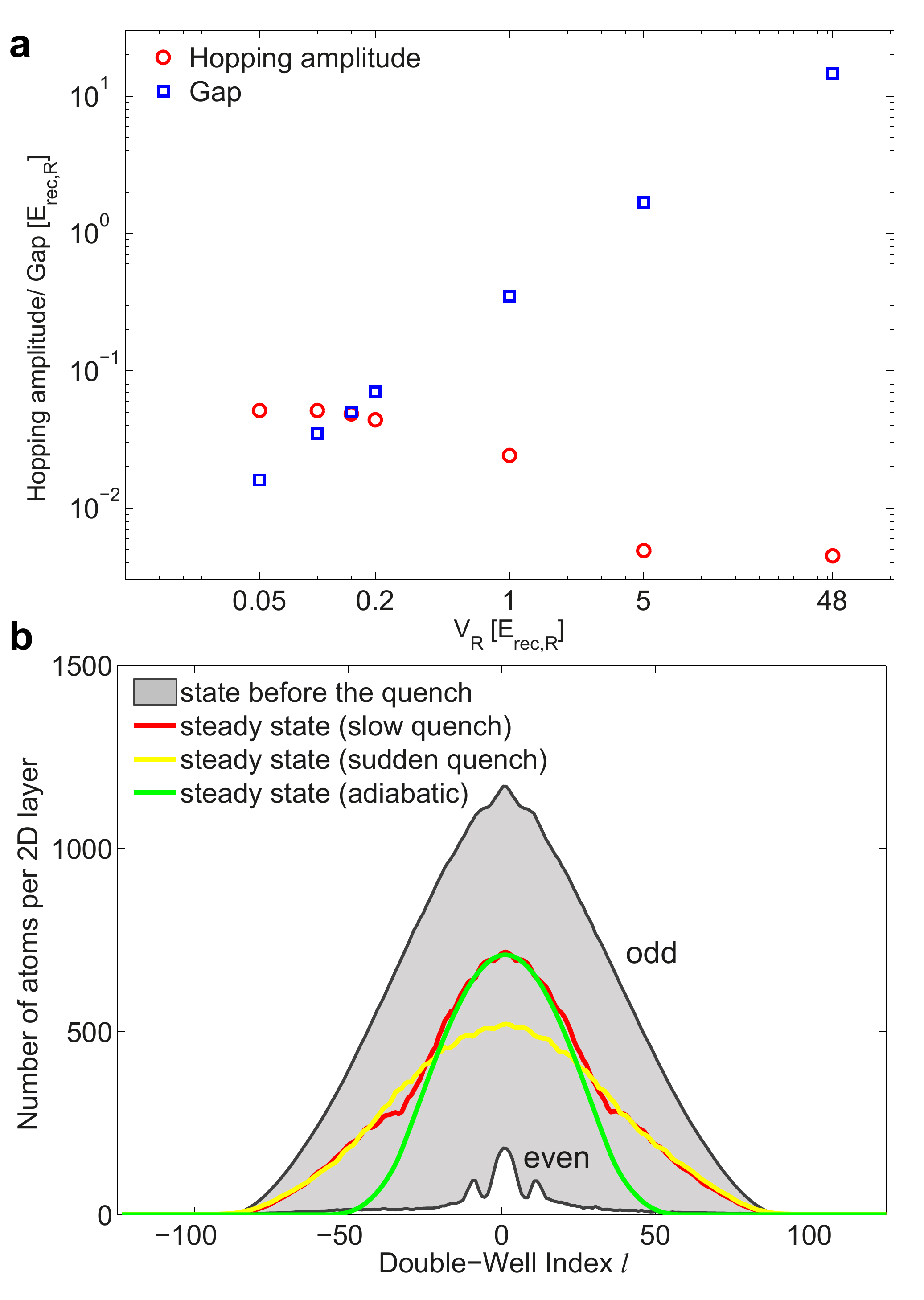}
 \caption{(Color online) (a) Hopping amplitude and local gap of the corresponding homogeneous system versus the depth of the red lattice potential. (b) Theoretical density profiles for $V_{R,f}=0$. Gray lines show the densities $n_{2l+1}$ and $n_{2l}$ versus double-well index $l$ before the slow quench. The colored lines show the stationary state after a slow ($\tau_{quench}=6$\,ms), a sudden ($\tau_{quench}=0.5\,$ms), and an adiabatic quench (all for $V_{R,f}=0$).} 
\label{fig4}
\end{figure}

In Fig. 4b the theoretical density profiles just before  he quench  for  $V_{R,f}=0$, and for long times (stationary situation) are shown to yield further spatial information. In the initial density profiles before the quench, the strong imbalance of the occupation between even and odd sites is shown. During the slow ramp down this observed  density imbalance is reduced and in the following time evolution this local imbalance oscillates and decays to zero, so that locally we obtain an equilibrated density. The global stationary profile, in contrast, can be decomposed into two superimposed shapes which reflects the non-equilibrium nature of the state. In the center the density profile resembles the stationary state after an {\it adiabatic} quench, whereas at the boundaries the distribution resembles the stationary state after a {\it sudden} quench. This clearly signals that the boundary regions do not relax. For non-zero  final red lattice depths the relaxation dynamics is much less pronounced due to the existence of the local gap. Even the local occupation imbalance freezes to a larger stationary value than expected for the corresponding adiabatic or equilibrium states.

To summarize, in this work we have investigated the relaxation dynamics of non-interacting fermions after different quenches of a one-dimensional superlattice potential. We have found that in this integrable system, for some quenches, a relaxation towards a state, which locally resembles an equilibrium state, occurs. However, the global density distribution signals the strongly non-equilibrium nature of the situation. In the future an investigation of the influence of interaction on relaxation processes would be of interest. In particular, in the present situation the interaction would lead to a coupling of the different directions of the lattice thereby opening more relaxation channels.

The work has been supported by the Alexander-von-Humboldt Professorship, BCGS, DFG, {EPSRC} (EP/G029547/1, EP/J01494X/1), the Leverhulme Trust, the Royal Society, and the Wolfson Foundation.

\newpage

\section{Supplementary material}

\subsection{Implementation of the optical superlattice potential}
The superlattice is created by overlapping two retro-reflected Gaussian beams of wavelengths $\lambda_R=1064\text{ nm}$ and $\lambda_G\approx\lambda_R/2$. To a good approximation, the potential along the $x$-direction is given by the sum of two lattices of depths $V_R$ and $V_G$,
\begin{equation}
V(x)=V_G\cos^2 (k_G x - \phi) -V_R \cos^2 (k_R x),
\end{equation}
where $k_{G/R}=2\pi/\lambda_{G/R}$ and $\phi$ is the phase difference between the two lattices at the place of the atoms. In the transverse plane, the red lattice beam is almost circular with $1/e^2$-radii of  $\text{w}_{y,R}\times \text{w}_{z,R} = 120\mathrm{\mu m}\times  130\mathrm{\mu m}$, while the green lattice beam is elliptic with $\text{w}_{y,G}\times \text{w}_{z,G} = 190 \mathrm{\mu m}\times  42\mathrm{\mu m}$ in order to better match the shape of our dipole trap and reduce the power required for a given lattice depth.

By detuning the frequency $\nu_G$ of the green laser from exactly twice the frequency $\nu_R$  of the infrared laser, we can control the shape of the double wells. The phase $\phi$ depends on the actual frequencies as
\begin{equation}
\phi= \phi_0 + \frac{2 \pi d}{c}\delta \nu ,
\end{equation}
where $\delta \nu = \nu_G - 2 \nu_R$, $c$ is the speed of light in vacuum and $d\approx \text{50 cm}$ is the optical path length for the green light from the atoms to the retro-reflecting mirror. The quantity $\phi_0$, which depends on the difference in refractive indices for the two different frequencies, can be considered constant during the time scale of the experiment. The detuning $\delta \nu$  is stabilized and adjusted by offset-locking the green laser to the frequency-doubled infrared laser light.

\subsection{Calibration of lattice depths and phase difference}

The lattice depths $V_G$ and $V_R$ are individually calibrated via modulation of the respective lattice depth. This induces a population transfer from the ground to the second excited band, which can be probed via band mapping. For the red lattice, we perform the calibration in the deep lattice regime, where the bands are essentially flat, and the resonance frequency is well defined. For the green lattice, we do not reach this regime due to limited available laser power and hence the inter-band transfer for a given modulation frequency depends on the momentum within the band. We therefore fit the momentum-resolved resonance frequency distribution \cite{Heinze2011} with the theoretical result derived from a band-structure calculation. The uncertainty of our calibration is 2\% for the red lattice and 8\% for the green lattice.

Unlike in the bosonic case \cite{Sebbystrabley2006,Sebbystrabley2007,Foelling2007}, for fermions the detuning that corresponds to a symmetric double well cannot be easily deduced from interference patterns after time of flight. We calibrate the phase difference $\phi$ as a function of the detuning $\delta \nu$ by observing the width of the atomic distribution along the direction of the superlattice after time of flight following a sudden turn-off of the lattice and trapping potentials. The wavefunction within a single well has the largest extent for the symmetric double-well configuration ($\phi=n \pi$), corresponding to a sharp minimum in the momentum width, that can be clearly observed in time-of-flight data for the appropriate ratio $V_G/V_R$. From the frequency distance between two minima the optical path length $d$ can be inferred. With $d$ and the detuning for the symmetric configuration, $\phi_0$ can be deduced up to a multiple of $\pi$.

\subsection{Detection of the local particle imbalance}
The detection of the local particle number in the even and odd sites of the green lattice is measured using a mapping onto different Brillouin zones. For this mapping we  quickly ramp up the red lattice to a depth of $V_R=48 E_{rec,R}$, which produces a highly imbalanced double-well (Fig.~1a, V) due to the prepared phase shift of $\phi = 0.48(2) \pi$. In doing so we project the population of each green lattice site onto the asymmetric superlattice sites. By subsequently ramping down the green lattice in 0.5\,ms we transfer the atoms from the odd (even) superlattice wells into the first (third and fourth)  Bloch bands of the red lattice (Fig. 1a, VI). The measurement sequence is completed by an adiabatic ramp down of the red lattice in 6\,ms (band-mapping)  in order to allow for a detection of the band populations and hence the (trap-averaged) populations of the even and odd lattice wells.

\subsection{Theoretical simulation}

 %+ \left. \left(\right)\Psi({\bf r})^\dagger\Psi({\bf r})
%\left(\frac{\partial^2}{\partial x^2}+\frac{\partial^2}{\partial y^2}+\frac{\partial^2}{\partial z^2}

The non-interacting fermionic atoms in the superlattice potential along $x$--direction and in the harmonic confinements along the $x$--, $y$-- and $z$--directions are described by the Hamiltonian
\begin{widetext}
\begin{eqnarray}
H&=&\int {\rm d}{\bf r}\left[\Psi^\dagger({\bf r}) \left(- \frac{\hbar^2}{2m}\nabla ^2
+V(x,t)+\frac{m}{2}\left(\omega_{x,0}^2x^2+ \omega_y^2(t)y^2+ \omega_z^2(t)z^2\right)\right)
\Psi({\bf r})\right],
\label{eq:Hxyz_time_sec_quant}
\end{eqnarray}
\end{widetext}
Here $\Psi({\bf r},t)$ and $\Psi^\dagger ({\bf r},t)$ are the fermionic field operators. The optical lattice potential along the $x$--direction is described by 
\begin{eqnarray}
V(x,t)=V_G(t)\cos^2(k_Gx -\phi)-V_R(t) \cos^2(k_Rx),
\end{eqnarray}
where the subindex $R,G$ denotes the red and green lattice, respectively. During the ramp  the intensity $V_\gamma$ follows the functional form
\begin{eqnarray}
&V_\gamma(t)=V_{\gamma,i}+(V_{\gamma,f}-V_{\gamma,i})\sin^2\left(\frac{\pi(t-t_{\gamma,0})}{2\tau_\gamma}\right),\nonumber
\end{eqnarray}
where $t$ is the time, $\gamma$ can be $R$ or $G$ for red and green lattice. The times $t_{\gamma,0}$ and $\tau_\gamma$ are the start time of the ramp and the ramp duration for the corresponding lattice and the given form $V_\gamma(t)$ is valid for times $t_{\gamma,0}\leq t \leq \tau_\gamma+t_{\gamma,0} $. $V_{\gamma,i}$ and $V_{\gamma,f}$ are the initial and final lattice depths. The dipole trapping frequencies are $\omega_{x,0}=2\pi\times 38$ Hz, $\omega_{y,0}=2\pi\times 66$ Hz, and $\omega_{z,0}=2\pi\times 376$ Hz. The trap frequencies in the $y$-- and the $z$--directions change during the ramp up of the red and green lattice due to the radial intensity profile of the beams. To a good approximation, for the red lattice beam the additional confinement is given by
\begin{eqnarray}
\omega_{\alpha,R}(t)=\left(\frac{4 V_R(t)}{m \text{w}_{\alpha,R}^2}\right)^{1/2}
\end{eqnarray}
and for the green lattice beam the deconfinement is given by
\begin{eqnarray}
\omega_{\alpha,G}(t)=\frac{h}{m\text{w}_{\alpha,G}\lambda_G}\left(\frac{V_G(t)}{E_{r,G}}\right)^{1/4}.
\end{eqnarray}
Here $\alpha$ denotes the $y$-- or $z$--directions and $\text{w}_{\alpha,\gamma}$ is the waist of the corresponding lattice beam. The frequency change in the $x$--direction is negligible during the ramping process. The resulting time dependencies of the trap frequencies are 
\begin{eqnarray}
&\omega_{\alpha}(t)=\left[\omega_{\alpha,0}^2+\omega_{\alpha,R}^2(t)-\omega_{\alpha,G}^2(t)\right]^{1/2},\quad\alpha=y,z\nonumber\\
&\omega_x(t)=\omega_{x,0}.
\label{eq:omega_time}
\end{eqnarray}

We diagonalize the Hamiltonian numerically using the fact that the Hamiltonian is quadratic in the fermionic operators. To simplify the calculation further we use the separability of the Hamiltonian in different directions, i.e.~$H=H_x+H_y+H_z$. The simulations are performed using exact diagonalization on the discretized version of the Hamiltonian and determinining the full time evolution of the fermionic gas during the experimental sequence. The space is discretized in the $x$--direction using $\Delta x=\frac{\lambda_{R}}{50}$ and along the $y$-- and $z$--directions the discretization used is $\Delta y=\Delta z=\frac{\lambda_{R}}{6}$. We find that these discretization parameters already give good results and only slight deviations as, for example, of an error of $5\%$ in the tunneling time scale  compared to a smaller discretization. 
The very slow loading procedure is performed with a time-step of 0.5 ms which mainly causes slight deviations in the beginning of the loading. Due to the experimental discreteness of the lattice ramps of about 0.1ms, we have chosen a corresponding time-step of 0.1 ms for the different lattice ramp downs.

\subsection{One-dimensional tight-binding model}
In addition, we use a tight-binding model at a filling $N=\frac{L}{2}$ to study the dynamics of fermions in one dimension after a quench (see also \cite{Flesch2008,Barmettler2009} for the bosonic case). The Hamiltonian before the quench is
\begin{eqnarray}
H=\sum_{j=1}^L\left[-J c_j^\dagger c_{j+1} -J c_{j+1}^\dagger c_j  + (-1)^j \frac{\Delta}{2}c_j^\dagger c_j\right]
\end{eqnarray}
where $J$ is the tunnelling amplitude, $L$ is the number of wells and $\Delta$ is the energy offset between neighbouring sites. The dimerization of the potential opens a gap of $\Delta$ in the energy band structure at half of the Brillouin zone of the green lattice. We prepare the system in the ground state corresponding to a certain value of $\Delta_0$. A sudden quench of the energy offset $\Delta_0$ to zero is performed and the system evolves after this quench. The  local density $n_j=\langle c_j^\dagger c_j\rangle$ can be calculated analytically as
\begin{eqnarray}
n_j(t)=\frac{1}{2}-(-1)^j\frac{1}{L}\sum_{k=1}^{L/2}\tanh(\theta_k) \cos\left[\frac{4Jt}{\hbar}\cos\left(\frac{2\pi k}{L}\right)\right],\nonumber
\end{eqnarray}
where $\sinh(\theta_k)=\frac{\Delta_0}{4J|\cos\left(\frac{2\pi k}{L}\right)|}$.
The evolution of the odd-even contrast is
\begin{eqnarray}
C(L,\Delta_0,t)=\frac{2}{L}\sum_{l=1}^{L/2}\left(n_{2l+1}(t)-n_{2l}(t)\right)\nonumber\\
=\frac{2}{L}\sum_{k=1}^{L/2}\tanh(\theta_k) \cos\left[\frac{4Jt}{\hbar}\cos\left(\frac{2\pi k}{L}\right)\right].
\end{eqnarray}
The dependence on  $\Delta_0$ is contained in $\theta_k$.
In the limit of $\frac{\Delta_0}{J}\to \infty$ only odd sites are occupied initially. The evolution in an infinite size system is then given by
\begin{eqnarray}
C(t)=J_0\left(\frac{4Jt}{\hbar}\right),
\end{eqnarray}
where $J_0$ is the zeroth-order Bessel function of the first kind. In this limit the averaged odd-even contrast is maximal in the beginning $C(0)=1$, then relaxes towards $C=0$ with oscillations of  period $T=\frac{h}{4J}$. When $\frac{\Delta_0}{J}$ is finite, the initial contrast is less than one and the following evolution shows again an oscillatory relaxation to $C=0$. %The oscillations seem to occur with the same frequency.

\end{document}